\documentstyle[sprocl,epsf]{article}
\def\Journal#1#2#3#4{{#1} {\bf #2}, #3 (#4)}

\def\NPB{{\em Nucl. Phys.} B}

\def\PRD{{\em Phys. Rev.} D}

\def\be{\begin{equation}}
\def\ee{\end{equation}}
\def\bea{\begin{eqnarray}}
\def\eea{\end{eqnarray}}

\def\gp{g^\prime}
\def\gps{g^{\prime^2}}
\def\gpp{g^{\prime\prime}}
\def\gpps{g^{\prime\prime^2}}

\def\xw{x^{\mbox{}}_W}
\def\xws{\mbox{\footnotesize$x^{\mbox{}}_{\mbox{\tiny$W$}}$}}
\def\mQ{m^{\mbox{}}_{q^\prime}}

\def\mheavyl{m^{\mbox{}}_{L}}

\def\tw{\theta^{\mbox{}}_W}

\def\mw{m^{\mbox{}}_W}
\def\mz{m^{\mbox{}}_Z}
\def\ms{m^{\mbox{}}_S}

\def\mhone{m^{\mbox{}}_{H^0_1}}
\def\ghone{\Gamma^{\mbox{}}_{H^0_1}}
\def\mhtwo{m^{\mbox{}}_{H^0_2}}
\def\ghtwo{\Gamma^{\mbox{}}_{H^0_2}}
\def\mhthree{m^{\mbox{}}_{H^0_3}}
\def\ghthree{\Gamma^{\mbox{}}_{H^0_3}}
\def\mzone{m^{\mbox{}}_{Z_1}}
\def\mztwo{m^{\mbox{}}_{Z_2}}
\def\gztwo{\Gamma^{\mbox{}}_{Z_2}}
\def\mpzero{m^{\mbox{}}_{P^0}}
\def\gpzero{\Gamma^{\mbox{}}_{P^0}}
\def\mhpm{m^{\mbox{}}_{H^\pm}}

\newcommand{\order}[1]{{\cal O}({#1})}
\def\teva{\mbox{T{\sc evatron}}}

\def\R{{\rm R}}
\def\L{{\rm L}}

\def\bra{\langle}
\def\ket{\rangle}

\newcommand{\Doublet}[2]{
   \left(\begin{array}{@{}c@{}}{#1}\\{#2}\end{array}\right)
}

   \def\NE{\nu_e}
   \def\Ep{e^\prime}
  \def\NpE{\nu_e^\prime}

  \def\NMU{\nu_\mu}
  \def\MUp{\mu^\prime}
 \def\NpMU{\nu_\mu^\prime}

 \def\NTAU{\nu_\tau}
 \def\TAUp{\tau^\prime}
\def\NpTAU{\nu_\tau^\prime}

   \def\UL{\left(\begin{array}{@{}c@{}}  
   u\\       d\end{array}\right)_\L}
   \def\EL{\left(\begin{array}{@{}c@{}} 
  \NE\\       e\end{array}\right)_\L}
  \def\EpL{\left(\begin{array}{@{}c@{}} 
 \NpE\\     \Ep\end{array}\right)_\L}
  \def\EpR{\left(\begin{array}{@{}c@{}} 
  \Ep\\    \NpE\end{array}\right)_\L^c}
				       
   \def\CL{\left(\begin{array}{@{}c@{}} 
     c\\      s\end{array}\right)_\L}
  \def\MUL{\left(\begin{array}{@{}c@{}} 
  \NMU\\    \mu\end{array}\right)_\L}
 \def\MUpL{\left(\begin{array}{@{}c@{}} 
 \NpMU\\   \MUp\end{array}\right)_\L}
 \def\MUpR{\left(\begin{array}{@{}c@{}} 
  \MUp\\  \NpMU\end{array}\right)_\L^c}
				       
   \def\TL{\left(\begin{array}{@{}c@{}} 
     t\\      b\end{array}\right)_\L}
 \def\TAUL{\left(\begin{array}{@{}c@{}} 
 \NTAU\\   \tau\end{array}\right)_\L}
\def\TAUpL{\left(\begin{array}{@{}c@{}}
 \NpTAU\\  \TAUp\end{array}\right)_\L}
\def\TAUpR{\left(\begin{array}{@{}c@{}} 
 \TAUp\\ \NpTAU\end{array}\right)_\L^c}

 \def\UR{u_\L^c}
 \def\DR{d_\L^c}
\def\NER{\nu_{e_\L}^c}
 \def\ER{e_\L^c}
\def\DpL{d_\L^\prime}
\def\DpR{d_\L^{\prime c}}
\def\NppE{\nu_{e_\L}^{\prime\prime c}}
	   
  \def\CR{c_\L^c}
  \def\SR{s_\L^c}
\def\NMUR{\nu_{\mu_\L}^c}
 \def\MUR{\mu_\L^c}
 \def\SpL{s_\L^\prime}
 \def\SpR{s_\L^{\prime c}}
\def\NppMU{\nu_{\mu_\L}^{\prime\prime c}}
	   
   \def\TR{t_\L^c}
   \def\BR{b_\L^c}
\def\NTAUR{\nu_{\tau_\L}^c}
 \def\TAUR{\tau_\L^c}
  \def\BpL{b_\L^\prime}
  \def\BpR{b_\L^{\prime c}}
\def\NppTAU{\nu_{\tau_\L}^{\prime\prime c}}
	 
\newcommand{\SU}[2]{\mbox{${\rm SU}({#1})_{{\rm #2}}$}}
\newcommand{\U}[2]{\mbox{${\rm U}({#1})_{{\rm #2}}$}}

\newcommand{\Esix}{\mbox{${\rm E}_6$}\ }
\newcommand{\ExE}{\mbox{${\rm E}_8\otimes{\rm E}_8^\prime\;$}}

\def\onetwo{\frac{1}{2}}

\newcommand{\ffrac}[2]{\mbox{\footnotesize$\frac{{#1}}{{#2}}$}}

\def\approxle{\,\raisebox{-0.625ex}{$\stackrel{<}{\sim}$}\,}
\def\approxge{\,\raisebox{-0.625ex}{$\stackrel{>}{\sim}$}\,}

\def\ggqzll{\raisebox{-1.26975cm}{
\mbox{
\setlength{\unitlength}{1cm}
\begin{picture}(5.743,3.6)
 \put(-0.257,0.086){
  \mbox{\epsfxsize=4.5cm
   \epsffile{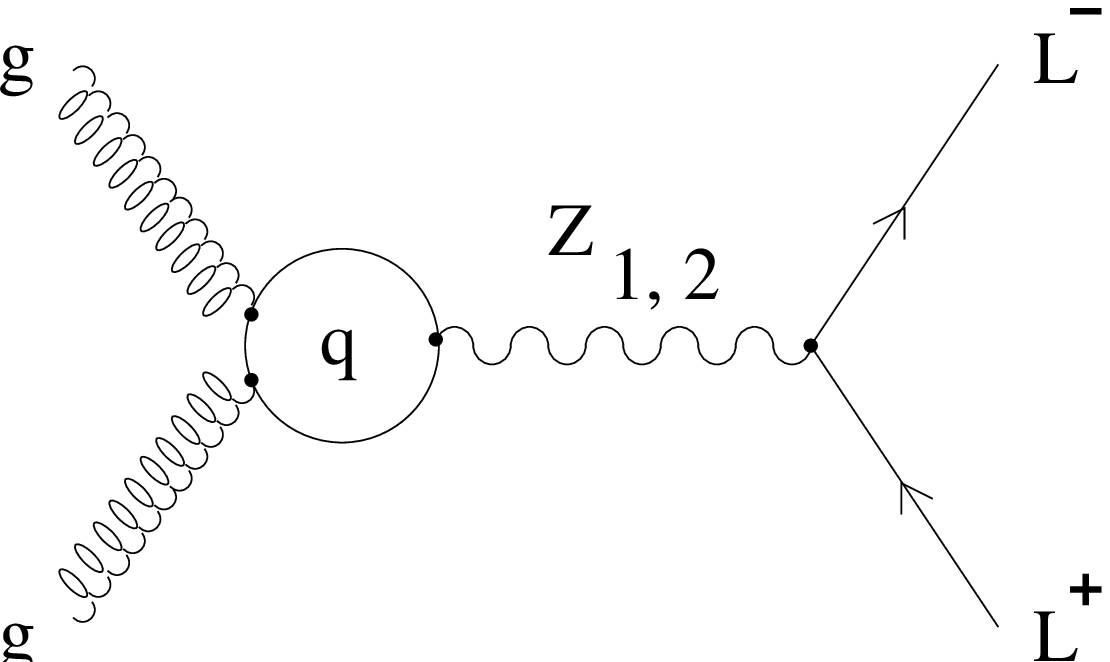}}
 }
\end{picture}
}}}

\def\ggqhpll{\raisebox{-1.26975cm}{
\mbox{
\setlength{\unitlength}{1cm}
\begin{picture}(4.30725,2.7)
 \put(-0.19275,0.0645){
  \mbox{\epsfxsize=4.5cm
   \epsffile{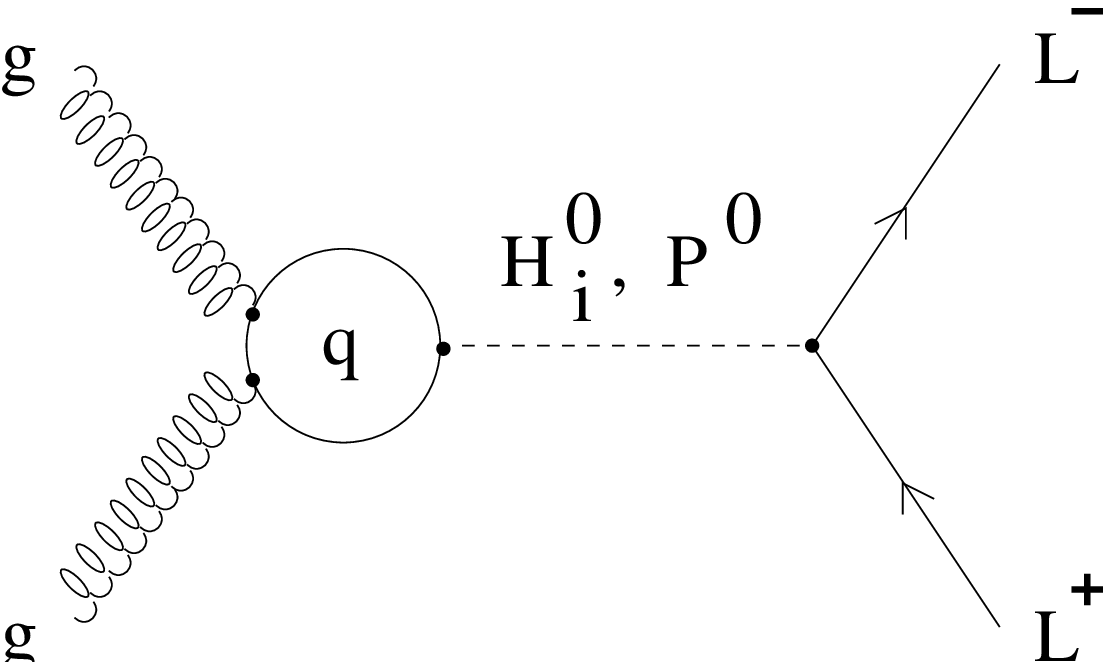}}
 }
\end{picture}
}}}

\def\ggsqhll{\raisebox{-1.26975cm}{
\mbox{
\setlength{\unitlength}{1cm}
\begin{picture}(4.30725,2.7)
 \put(-0.19275,0.0645){
  \mbox{\epsfxsize=4.5cm
   \epsffile{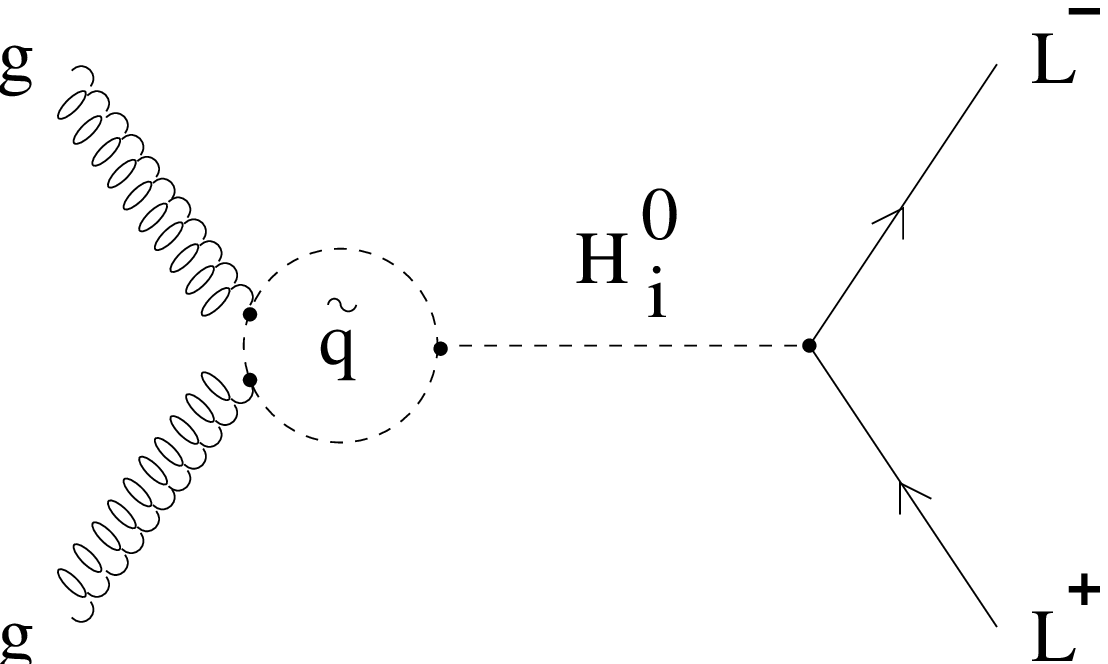}}
 }
\end{picture}
}}}

\def\ggvsqhll{\raisebox{-1.26975cm}{
\mbox{
\setlength{\unitlength}{1cm}
\begin{picture}(4.30725,2.7)
 \put(-0.19275,0.0645){
  \mbox{\epsfxsize=4.5cm
   \epsffile{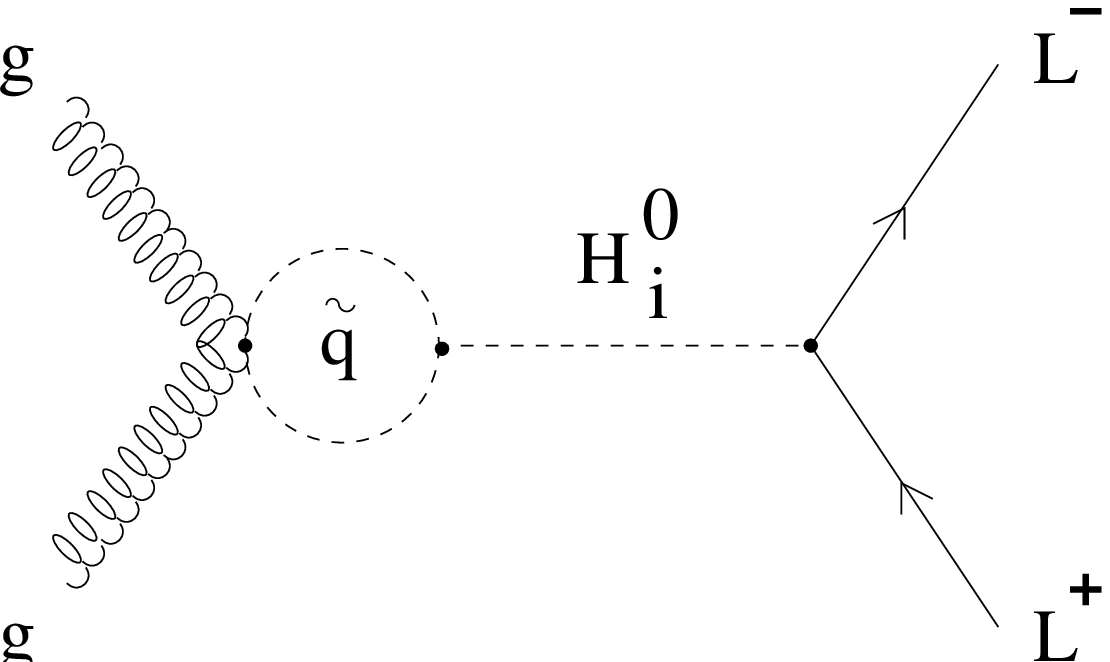}}
 }
\end{picture}
}}}

\begin{document}
\mbox{}\hfill OCIP/C-96-15\\
\mbox{}\hfill UQAM-PHE-96-01\\
\title{HEAVY CHARGED    LEPTON PRODUCTION   IN SUPERSTRING  INSPIRED  E6
MODELS\cite{kn:Boyce,kn:BMK}} 
\author{M.M.~BOYCE,\footnote{speaker      (copies      available     at,
http://www.physics.carleton.ca/~boyce/)} M.A.~DONCHESKI}
\address{Carleton University, Physics Department, 1125 Colonel By Drive,
Ottawa,\\   Ontario,   K1S  5B6,   Canada}
\author{ H.~K\"{O}NIG  }  \address{UQAM, D\'epartement   de Physique, CP
8888, Succ.  Centre Ville, Montr\'eal,\\ Qu\'ebec, H3C 3P8, Canada}
%
\maketitle\abstracts{ We  investigate the possibility  of studying E$_6$
phenomenology at high energy  hadron colliders.  The production of heavy
leptons pairs {\it via} a gluon-gluon fusion mechanism is discussed.  An
enhancement in  the parton  level cross-section  is expected  due to the
heavy (s)fermion loops which couple to the gluons.}
\section{Introduction}
    In  this  talk we  give  an overview  of   a very simple superstring
inspired \Esix model followed by an  application to heavy charged lepton
production, along with some preliminary results and discussion.

\subsection{Our Choice of Model}
    The interest  in  \Esix  as  a $GUT$,   from   its good old   1970's
would-be-``topless'' days, was  rekindled in  late  1984 when Green  and
Schwarz showed that  the \ExE ga\-uge group  gives rise  to anomaly free
10-d   string theory.   In  particular, the  Calabi-Yau compactification
scheme   $$\ExE\;\longrightarrow\;{\rm SU(3)}\otimes{\rm E}_6\otimes{\rm
E}_8^\prime\;,$$ down  to 4-d (assuming N=1 SUSY)\cite{kn:Hewett} yields
$n_g$  copies of the {\bf 27}  representation of \Esix, depending on the
topology  of  the compactified space.   The  ${\rm SU(3)}$ are  the spin
connections on  the   compactified  space, and  the   ${\rm E}_8^\prime$
couples by gravitational interactions  to the matter representations  of
E$_6\,$; these interactions should presumably play a role in lifting the
supersymmetric mass degeneracy.  It should  be pointed out that \Esix is
by no means a unique $GUT$ group, as other dimensional reduction schemes
can give rise  to different  ones.  However,  a lot of  phenomenological
work  has  be done  using   \Esix and therefore  we  shall   humbly take
advantage of this fact.

  Further breaking of \Esix, {\it via}  the Hosotani mechanism, leads to
rank-6 and rank-5 groups.  Here the simplest  of these scenarios will be
considered,       {\it      i.e.},     $$\Esix           \longrightarrow
\underbrace{\SU{3}{c}\otimes\SU{2}{L}\otimes\U{1}{Y}}_{\mbox{   Standard
Model  ($SM$)}}\otimes\U{1}{Y_E}\,,$$ where the extra $\U{1}{Y_E}$ leads
to an additional neutral  gauge boson, the $Z^\prime$,  to the $SM$.  To
obtain $SM$  energies the breaking can now  proceed by  the conventional
Higgs mechanism.

  The assignment of the quantum numbers to the matter fields in the {\bf
27}, with  the exception of the $SM$  content, can  lead to models which
contain leptoquarks, diquarks, or  quarks.   Here, the  assignment  that
leads  to  the   more   familiar  $SM$-like  quarks   will   be  chosen,
Figure~\ref{fig:esix}.
\begin{figure}[htb]
$${\bf 27}=\left\{\fbox{$\UL\EL\UR\DR\ER$}\;
  \NER\DpL\DpR\EpL\EpR\NppE\right\}$$
$${\bf 27}=\left\{\fbox{$\;\CL\MUL\CR\SR\MUR$}\;
  \NMUR\SpL\SpR\MUpL\MUpR\!\!\NppMU\right\}$$
$${\bf 27}=\left\{\fbox{$\;\;\TL\TAUL\TR\BR\TAUR$}\;
  \NTAUR\BpL\BpR\TAUpL\TAUpR\NppTAU\right\}$$
$$Y_E^{\mbox{}}\{{\bf 27}\}=\left\{
  \fbox{$\displaystyle
     \;\;\left(\frac{2}{3}\right)\left(\frac{-1}{\;\;3}\right)\;
     \!\ffrac{2}{3}\;\ffrac{-1}{\;\;3}\;\ffrac{2}{3}
  $}\;\ffrac{5}{3}\;\ffrac{-4}{\;\;3}\;\ffrac{-1}{\;\;3}
  \left(\frac{-1}{\;\;3}\right)\left(\frac{-4}{\;\;3}\right)
  \;\;\ffrac{5}{3}\;
\right\}
\;\;\;\;\;\;\;\;
$$
\caption{\Esix  particle  content.  On  the   top  three  rows,  the  SM
   particles  are shown in the boxes   on the left  and their ``exotic''
   counter parts outside the boxes on the  right.  The exotics have been
   labeled  such  that they  carry the ``expected''  SM quantum numbers,
   with the  exception being L=0 for the  primed and double primed ones.
   The      bottom   row     contains    the  $Y_E^{\mbox{}}$    quantum
   numbers.\hfill\mbox{}}
\label{fig:esix}
\end{figure}
In this model,  the third generation   sleptons are typically  chosen to
play the role of the Higgs fields,
$$
\Phi_1=
  \Doublet{\tilde\nu^\prime_{\tau_\L^{\mbox{}}}}{
     \tilde\tau^\prime_{\scriptscriptstyle\L}}
  \equiv\Doublet{\phi_1^0}{\phi_1^-}\,,\;
\Phi_2=
  \Doublet{\tilde\tau^{\prime^c}_{\scriptscriptstyle\L}}{
     \tilde\nu^{\prime^c}_{\tau_\L^{\mbox{}}}}
  \equiv\Doublet{\phi_2^+}{\phi_2^0}\,,\;
\Phi_3=\tilde\nu^{\prime\prime^c}_{\tau_\L^{\mbox{}}}\equiv\phi_3^0\,,
$$
where    $\Phi_k=(\phi^a_{kR}+i\phi^a_{kI})/\sqrt{2}\,$, with    $VEV$'s
$\bra\phi_{kR}^0\ket=v_k\,$.
Therefore, the most general superpotential is of the form
$$
\begin{array}{@{}l@{\,}c@{\,}l@{}}
W&\sim&
 \lambda_1(u\;d)_\L^{\mbox{}} i\tau_2\EpR\UR
 +\lambda_2(\NpE\;\Ep)_\L^{\mbox{}} i\tau_2\UL\DR
 +\lambda_3(\NpE\;\Ep)_\L^{\mbox{}} i\tau_2\UL\ER\\&&\\&&
 +\lambda_4(\NpE\;\Ep)_\L^c         i\tau_2\EpL\NppE
 +\lambda_5\DpL\DpR\NppE+\ldots\,,
\end{array}
$$
where  the generation indices on   the Yukawa couplings,  $\lambda_a\,$,
have been  suppressed  ({\it   i.e.}, $\lambda_a\-\sim\-\lambda_a^{ijk}$
s.t.            $i,j,k=1,2,3$),                such                 that
$\lambda_4^{i33}=\-\lambda_4^{i33}=\-\lambda_4^{i33}=0$ for $i=1,2\,$.
$W$, and the aforementioned gauge  group, specifies all of the couplings
in this supersymmetric model,
$$
\begin{array}{l@{\,}c@{\,}l@{\;\;\;\;}l}
{\cal L}_{\rm Yuk}^{\mbox{}}&=&-\frac{1}{2}\left[
\frac{\partial^2W}{\partial\tilde f_i\partial\tilde f_j}\,f_if_j
+\frac{\partial^2W}{\partial\tilde f_i\partial\tilde f_j}
\!\raisebox{1.25ex}{*}\bar f_i\bar f_j\right]&
\mbox{\underline{Yukawa}}\\
V_{\rm Scalar}^{\mbox{}}&=&\underbrace{
   \left|\frac{\partial W}{\partial\tilde f_i}\right|^2}_{V_F^{\mbox{}}}
+\underbrace{
   \ffrac{1}{2}(g\tilde f^*_iT^a_{ij}\tilde f_j)^2+\ffrac{1}{4}
   (\gp Y_i\tilde f^*_i\tilde f_i)^2+\ffrac{1}{4}
   (\gpp Y_{E_i}\tilde f^*_i\tilde f_i)^2}_{V_D^{\mbox{}}}&
\mbox{\underline{Scalar}}
\end{array}
$$
where $g$, $\gp$,  and $\gpp$    ($\approx\gp$) are  the    $\SU{2}{L}$,
$\U{1}{Y}$, and $\U{1}{Y_E}$ coupling constants, respectively.  In order
to lift the  supersymmetric degeneracy some soft terms  are put into the
model by hand,
$$
V_{\rm Soft}^{\mbox{}}\supseteq\tilde M^2_{\rm Q}
\left|\Doublet{\tilde u_\L^{\mbox{}}}{\tilde d_\R^{\mbox{}}}\right|^2+
\tilde M^2_u\tilde u_\R^{\mbox{}}\tilde u_\R^*+\ldots
+\lambda_1A_u(\tilde u\;\tilde d)_\L^{\mbox{}}i\tau_2
\underbrace{
   \Doublet{\tilde\tau^{\prime^c}_{\scriptscriptstyle\L}}{
        \tilde\nu^{\prime^c}_{\tau_\L^{\mbox{}}}}}_{\rm Higgses}
   \tilde u_\R^*
+\ldots\,,
$$
with the soft-SUSY breaking terms $\tilde M^2_f$ and  $A_f\,$.

   Now,   the  Higgs potential,    $V_H^{\mbox{}}$  ($\subseteq   V_{\rm
Scalar}^{\mbox{}}+V_{\rm Soft}^{\mbox{}}$), produces  masses  for all of
the particles  in   the model, with  its  remaining  degrees of  freedom
creating    physical     Higgs     fields;     three    neutral-scalars,
$H^0_{i=1,2,3}\,$,    a      neutral-pseudo-scalar,  $P^0\,$,    and   a
charged-scalar, $H^\pm\,$.    The  mass mixing matrices  for   the Higgs
fields, ${\cal M}_{\rm  Higgs}^2\,$, can be  obtained from  the bilinear
terms in $V_H$,
$$
V_H(\phi_k^a)\;\supseteq\;\onetwo\,{\cal M}_{ij}^2\,
(\phi_i^a-\langle\phi_i^a\rangle)(\phi_j^a-\langle\phi_j^a\rangle)
\;\;\ni\;\;
{\cal M}_{ij}^2=\left.
\frac{\partial^2V_H}{\partial\phi_i^a\partial\phi_j^a}
\right|_{VEV's}\,,
$$
which   produces a $(2\times2)$    for ${\cal  M}_{H^\pm}^2\,$, and  two
$(3\times3)$'s        for  ${\cal       M}_{H^0_i}^2$    and      ${\cal
M}_{P^0}^2\,$. Diagonalizing these matrices yields the mass terms
$$
m^2_{H^\pm}=\frac{\lambda A\nu_3}{\sin(2\beta)}
+\left(1-2\,\frac{\lambda^2}{g^2}\right)m_W^2\,,\;\;
m_{P^0}^2=\frac{\lambda A\nu_3}{\sin(2\beta)}\,
\left(1+\frac{\nu^2}{4\nu_3^2}\sin^2(2\beta)\right)\,,
$$
and
$$
m^2_{H^0_i}\subseteq(U{\cal      M}_{H^0_i}^2U^{-1})_{D}^{\mbox{}}
\;\;\ni\;\;
m^2_{H^0_1}\le m^2_{H^0_2}\le m^2_{H^0_3}\,,
$$
where            $\tan\beta=v_2/v_1\,$,             $v^2=v_1^2+v_2^2\,$,
$\lambda=\lambda_a^{333}\,$,  and  $A$  is   a  soft  term.\footnote{The
$\tilde   M_f^2\in    V_H$ are    eliminated    by  requiring  $\partial
V_H/\partial\phi_i^a|_{\mbox{\tiny  $VEV$'s}}^{\mbox{}}=0\,$.}  Assuming
a $U$-gauge, the $\phi_i^a$'s can be  expressed in terms of the physical
Higgs fields, 
$$
\phi^\pm_1=\sin\beta\,H^\pm\,,\;\;\;\;\;\phi^\pm_2=\cos\beta\,H^\pm\,,
\;\;\;\;\;\;\;\;\;\;\;\;\;\;\;\;\;\;\;\;\;\;\;\;
\;\;\;\;\;\;\;
$$
$$
\phi^0_{1I}=\kappa\,v_2v_3\,P^0\,,\;\;\;
\phi^0_{2I}=\kappa\,v_1v_3\,P^0\,,\;\;\;
\phi^0_{3I}=\kappa\,v_1v_2\,P^0\,,
$$
and
$$
\phi^0_{iR}=\nu_i+\sum_{j=1}^3U_{ij}^{-1}H^0_j\,.
\;\;\;\;\;\;\;\;\;\;\;\;\;\;\;\;\;\;\;\;\;\;\;\;
\;\;\;\;\;\;\;\;\;\;\;\;\;\;\;\;\;\;\;\;\;\;\;\;\;
$$
where  $\kappa=1/\sqrt{v_1^2v_2^2+\nu^2\nu_3^2}\,$.  This allows for the
other masses, and  interactions   involving the Higgses,  to   be easily
obtained.

  The masses  for the gauge bosons come  from the kinetic terms  of the
Higgs fields, as follows,
$$
\begin{array}{@{}l@{\,}c@{\,}l@{}}
{\cal L}_{\rm K.E.}^{\Phi_i}&\supseteq&
|(\partial-\ffrac{i}{2}G_\mu)\Phi_i|^2\nonumber\\
&\supseteq&
m_W^2W^+_\mu W^{-\mu}+\frac{1}{2}(Z Z^\prime)_\mu
\underbrace{
  m_Z^2\left(\begin{array}{@{}cc@{}}
  1&\frac{4v_2^2-v_{1_{\mbox{}}}^2}{3v^2/\eta}\sqrt{\xws}\\
  \frac{4v_2^2-v_{1_{\mbox{}}}^2}{3v^2/\eta}\sqrt{\xws}&
  \frac{v_{1_{\mbox{}}}^2+16v_2^2+25v_3^2}{9v^2/\eta^2}\xws
  \end{array}\right)
}_{
  {\cal M}^2_{Z\mbox{-}Z^\prime}
}
\Doublet{Z}{Z^\prime}^\mu\!,\nonumber
\end{array}
$$
where
$$
G_\mu\supseteq(\tau_3g\cos\tw-\gp Y\sin\tw)Z_\mu
+\sqrt{2}g[\tau_+W^-_\mu+\tau_-W^+_\mu]
+\gpp Y_EZ_\mu\,,
$$
$\mw=gv/2\,$,      $\mz=\sqrt{g^2+\gps}\,v/2\,$,   and  $\eta=\gp/\gpp$.
Therefore, the $Z$ and $Z^\prime$ mix to give the mass eigenstates $Z_1$
and   $Z_2$,      with       masses     $m_{Z_1}^{\mbox{}}(\equiv\mz)\le
m_{Z_2}^{\mbox{}}\,$.

  The fermion   masses come from the Yukawa   interaction terms with the
Higgs fields,
$$
{\cal L}_{\rm Yuk}\supseteq
-\frac{1}{\sqrt{2}}\,\left\{
\lambda_1\nu_2\,\bar u u+
\lambda_2\nu_1\,\bar d d+
\lambda_3\nu_1\,\bar e e+
\lambda_4\nu_3\,\bar e^\prime e^\prime+
\lambda_5\nu_3\,\bar d^\prime d^\prime
\right\}\,,
$$
which yields the Yukawa couplings
$$
\lambda_1=\frac{g\,m_u}{\sqrt{2}\,\mw\sin\beta}\,,\;\;\;\;
\lambda_2=\frac{g\,m_d}{\sqrt{2}\,\mw\cos\beta}\,,\;\;\;\;
\lambda_3=\frac{g\,m_e}{\sqrt{2}\,\mw\cos\beta}\,,
$$
$$
\lambda_4=\frac{\sqrt{2}}{\nu_3}\,m_{e^\prime}\,,\;\;\;\;
\;\;\;\;\;\;\;\;\;
\lambda_5=\frac{\sqrt{2}}{\nu_3}\,m_{d^\prime}\,.
\;\;\;\;\;\;\;\;\;\;\;\;\;\;\;\;\;\;\;\;\;\;\;\;\;\;\;\;
\;\;\;\;\;\;\;\;\;\;\;\;\;\;\;\;\;
$$
 
  The   left   and    right  sfermion  states,    $\tilde f_{\mbox{\tiny
L,R}}^{\mbox{}}\,$, in general, mix and are obtained from
$$
V_{\rm Scalar}^{\mbox{}}+V_{\rm Soft}^{\mbox{}}\supseteq
(\tilde f_\L \tilde f_\R)
\underbrace{
  \left(\begin{array}{@{}cc@{}}
    {\cal M}_{\L\L}^2 & {\cal M}_{\L\R}^2\\
    {\cal M}_{\L\R}^2 & {\cal M}_{\R\R}^2
  \end{array}\right)
}_{
  {\cal M}_{\tilde f}^2
}
\Doublet{\tilde f_\L^*}{\tilde f_\R^*}
\;\;\longrightarrow\;\;
\sum_{i=1}^2 m_{\tilde f_i}^2\tilde f_i^{\mbox{}}\tilde f_i^*\,,
$$
where ${\cal M}_{\tilde f}^2$ is sfermion  mass mixing matrix: e.g.  for
$\tilde u_{\mbox{\tiny L,R}}^{\mbox{}}$
$$
M^{(\tilde u)^2}_{LL}=
\tilde M_Q^2+m_u^2+\ffrac{1}{6}\,(3-4\xw)\,m_Z^2\,\cos(2\beta)-
\ffrac{1}{36}\,\gpps(\nu_1^2+4\nu_2^2-5\nu_3^2)\,,
$$
$$
M^{(\tilde u)^2}_{RR}=
\tilde M_u^2+m_u^2+\ffrac{2}{3}\,\xw\,m_Z^2\,\cos(2\beta)-
\ffrac{1}{36}\,\gpps(\nu_1^2+4\nu_2^2-5\nu_3^2)\,,
\;\;\;\;\;\;\;\;\;\;\;\;
$$
and
$$
M^{(\tilde u)^2}_{LR}=m_u\,(A_u-m_{e^\prime}\cot\beta)\,.
\;\;\;\;\;\;\;\;\;\;\;\;\;\;\;\;\;\;\;\;\;\;\;\;\;\;\;\;
\;\;\;\;\;\;\;\;\;\;\;\;\;\;\;\;\;\;\;\;\;\;\;\;\;\;\;\;
\;\;\;\;\;\;\;\;\;\;
$$

   This concludes our survey of a simple E$_6$ model, as we now have all
of   the relevant phenomenology   for  discussing $L^+L^-$ production at
hadron colliders.\footnote{For more details  on the material covered  in
this section see \cite{kn:Boyce,kn:Hewett,kn:EllisB,kn:Haber}.}

\section{Heavy Lepton Production}
\begin{figure}[htb]
$$
    \mbox{
      \epsfxsize=5cm
      \epsffile{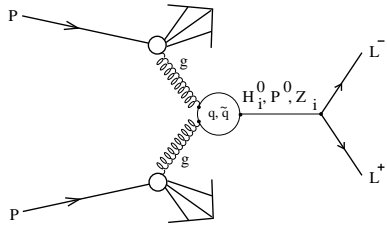}
    }
$$
\caption{$pp\rightarrow gg\rightarrow L^+L^-$}
\label{fg:cool}
\end{figure}

  An interesting place  to look for signatures  of new physics  is heavy
lepton production at high energy hadron colliders, figure~\ref{fg:cool}.
For        heavy    charged       leptons,    $L^\pm$,      with    mass
$\mheavyl\approxge\order{150}GeV$  the predominate production  mechanism
is gluon-gluon fusion,   within the $SM$\cite{kn:Yang} and  the  minimal
supersymmetric     standard  model ($MSSM$).\cite{kn:Montalvo}  For high
enough energies then,  the gluon luminosity of the  hadrons can be taken
advantage of. Accelerators that may provide good hunting grounds are
\begin{itemize}
\item \hspace{2em} $LHC\;\;\;\;\;\;\;\;\;$ \hspace{2em} ($pp$)  
    \hspace{2em} $\;14TeV$ \hspace{2em} ${\cal L}\sim10^5pb^{-1}/yr\,$,
\item \hspace{2em} $\teva$ \hspace{2em} ($p\bar p$)   
    \hspace{2em} $1.8TeV$ \hspace{2em} ${\cal L}\sim10^2pb^{-1}/yr\,$.
\end{itemize}
It     is   expected    that    the    parton    level    cross-section,
$\hat\sigma(gg\rightarrow L^+L^-)\,$,  should   be enhanced due   to the
number of heavy  particles running around  in the loop. This process has
been   computed   in   the    $MSSM$   by  Cieza   Montalvo,     {\it et
al.},\cite{kn:Montalvo}  which predicts  $\order{10^5}pb^{-1}/yr$    for
$50\le\mheavyl\le400GeV$.  Therefore,  since \Esix has more particles it
is  expected  that its $L^+L^-$   production  rate should  be  even more
enhanced.

\subsection{$gg\rightarrow L^+L^-$}
\begin{figure}[htb]
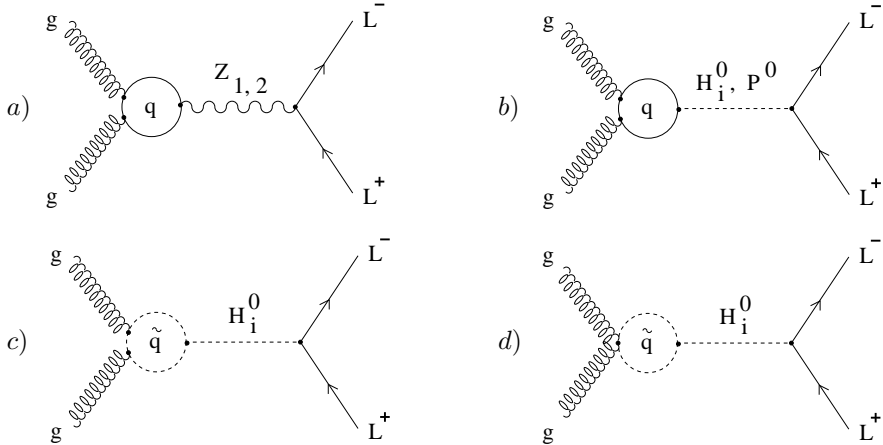

$$
\begin{array}{l@{}l@{}l@{}l}
  a) & \ggqzll  & b) &\ggqhpll  \\ &&& \\
  c) & \ggsqhll & d) &\ggvsqhll
\end{array}
$$
\caption{Feynman   diagrams for   gluon-gluon  fusion to  heavy  charged
   leptons.}
\label{fg:fusion}
\end{figure}

  Figure~\ref{fg:fusion}  show the Feynman  diagrams that  are needed to
compute   the  parton   level    cross-section  (matrix  elements)   for
$gg\rightarrow L^+L^-$.
These   matrix   elements are  similar   to   Cieza  Montalvo,  {\it  et
al.},\cite{kn:Montalvo} and  therefore  with some care  can be extracted
from their paper. The general form of the cross-section is,
$$
\hat\sigma=  \hat\sigma_{qZ_{1,2}}^{\mbox{}}
           + \hat\sigma_{qH_{1,2,3}^0}^{\mbox{}}
           + \hat\sigma_{qP^0}^{\mbox{}}
           + \hat\sigma_{\tilde qH_{1,2,3}^0}^{\mbox{}}
           +\underbrace{
                \hat\sigma_{q(Z_{1,2}-P^0)}^{\mbox{}}
              + \hat\sigma_{(\tilde q-q)H_{1,2,3}^0}^{\mbox{}}
           }_{
             {\em Interference Terms}
           }\,.
$$
We have decided  not to show the explicit details,\cite{kn:Boyce,kn:BMK}
since they are not  very enlightening.  However,  it worth pointing  out
that,  in the large $v_3$  limit only the  terms involving $Z_{1,2}$ and
$H^0_3$ survive.  Before the cross-section  can be computed some of  the
\Esix model parameters need to be constrained.

\begin{figure}[htbp]
\mbox{
\setlength{\unitlength}{1cm}
\begin{picture}(11.9,5.25)
 \put(-0.25,-0.5){\mbox{\epsfxsize=6.5cm
	\epsffile{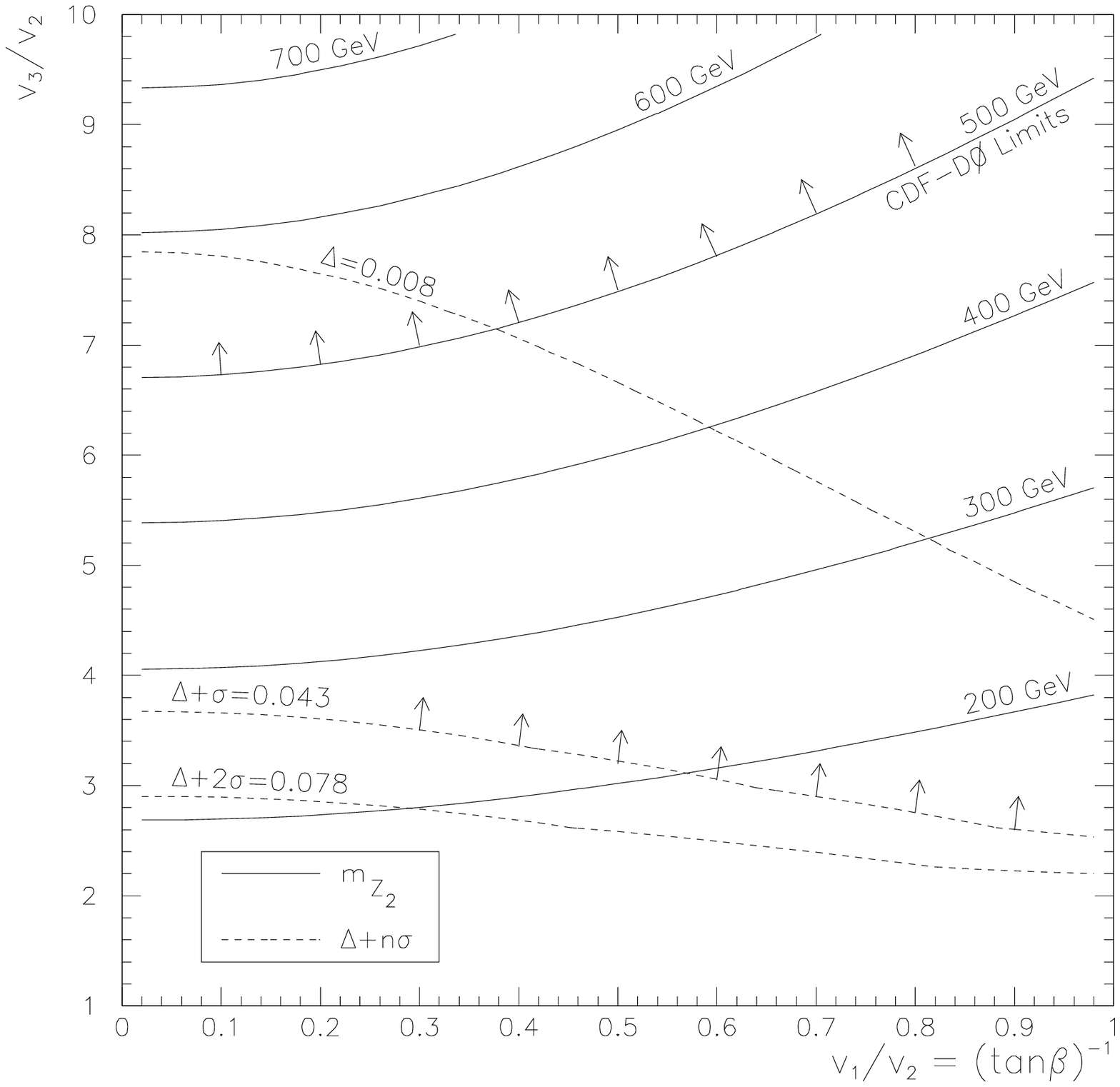}}}
 \put(5.65,-0.5){\mbox{\epsfxsize=6.5cm
	\epsffile{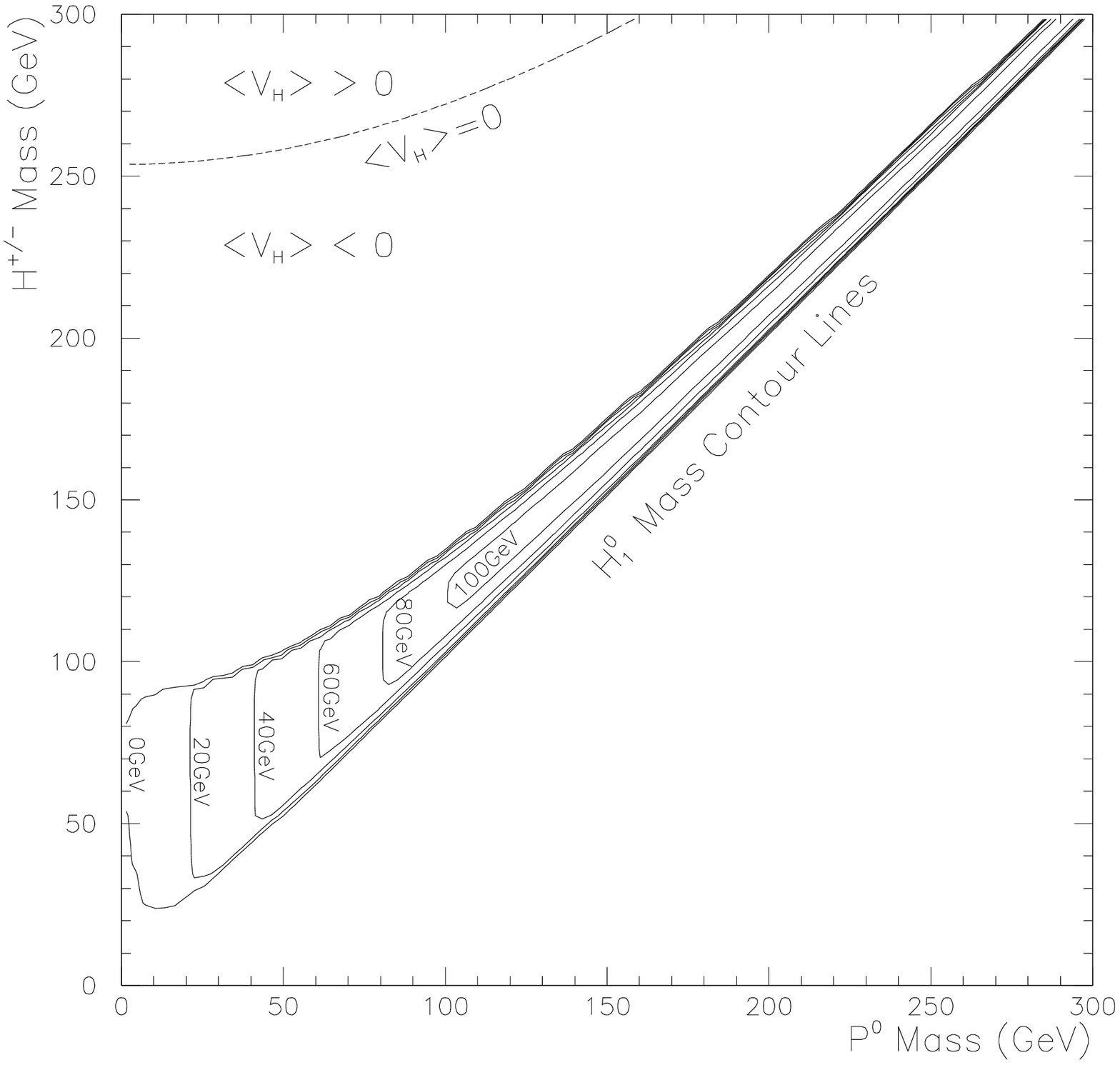}}}
 \put(2.75,0.5){\bf (a)}
 \put(8.75,0.5){\bf (b)}
\end{picture}}
\caption[foo]{Plots  of;  (a) $m_{Z_2}^{\mbox{}}$ and  $\Delta$  contour
     lines as  a function  of $v_1/v_2\,$  and  $v_3/v_2$.  The $\Delta$
     contour lines   are shown  at   the  $0\sigma\,$, $1\sigma\,$,  and
     $2\sigma$     levels        ($cf$~\cite{kn:EllisA}).            The
     $m_{Z_2}^{\mbox{}}=500GeV$   line shows the   $CDF$ and $D0\!\!\!/$
     soft-limits.\cite{kn:Shochet}   The   arrows point   to toward  the
     allowed regions on the plot.  (b) $m_{H^0_1}^{\mbox{}}\ge0$ contour
     lines as  a function of  $\mpzero$ and $\mhpm$,  for $v_1/v_2=0.02$
     and $v_3/v_2=6.7\,$.   The dashed   curve in the    upper left-hand
     corner is a plot of the zero of  the Higgs potential above which it
     becomes positive.\hfill\mbox{}}\label{fg:cnst}
\end{figure}
  The $\mzone$ ({\it  i.e.}, $\mz$)  mass can  be used  to constrain the
VEV's  by     requiring  it  lies    within  experimental   bounds, {\it
i.e.},\cite{kn:EllisA}
$$
\underbrace{\sin^2\bar\theta_W^{\mbox{}}}_{\;\;\;\;\bar\xw}\equiv 
1-\frac{m_W^2}{m_{Z_1}^2}\;\;<\;\;
\underbrace{\sin^2\theta_W^{\mbox{}}}_{\;\;\;\;\xw}\equiv 
\left.\frac{\gps}{g^2+\gps}\right|_{\mu=\mw}\,,
$$
due  to  mixing  with $\mztwo\,$.    Therefore,  to within  experimental
fluctuations  in $\bar\xw$ ($\approx 0.2247\pm0.0019$\cite{kn:PDG})  and
$\xw$ ($\approx 0.233\pm0.035|_{\mu=\mw}^{\mbox{}}$) we have
$$
\Delta\equiv\xw-\bar\xw\approx0.008\pm0.035\,.
$$
Figure~\ref{fg:cnst}.a shows the $\Delta$  and $\mztwo$ contour lines as
a function  of  $v_1/v_2$ ({\it   i.e.}, $1/\tan\beta$) and   $v_3/v_2$.
Here, $v_1/v_2\approxle\order{1}$ has been  chosen, since $m_b<<m_t$ for
any reasonable   range of  Yukawa   couplings.\cite{kn:EllisA,kn:Gunion}
Taking $\Delta+1\sigma$ contour line gives $v_3/v_2\approxge\order{3.5}$
($cf$ \cite{kn:Hewett,kn:EllisA}) or $\mztwo \approxge  \order{200}GeV$.
Since  the  error  in $\Delta$  is  fairly  large,  a  more conservative
approach has   been   taken by   invoking  the  $CDF$  and   $D0\!\!\!/$
soft-limits   ({\it    i.e.},  assuming    $SM$  couplings)  on $\mztwo$
(fig.~\ref{fg:cnst}.a).\cite{kn:Shochet} These give  fairly   reasonable
bounds  on  $\mztwo$,  {\it   i.e.}, $\mztwo  \approxge  \order{500}GeV$
($\Rightarrow$     $v_3/v_2\approxge\order{7.5})$,     since    $Y_E{\rm
's}\sim{\cal O}(Y)$'s ($cf$ fig.~\ref{fig:esix}).  For the moment let us
make the further assertion that $\lambda_1\sim\lambda_2\,$. Then we have
$v_1/v_2=0.02$     ({\it i.e.},  $\sim\order{m_b/m_t}$),   which implies
$v_3/v_2=6.7$ for $\mztwo\approx\order{500}GeV$.

   Now  that  the $VEV$'s  have  been fixed,  the Higgses  masses can be
constrained by adjusting $\lambda$   and $A$, or  equivalently $\mpzero$
and $\mhpm$.    Figure~\ref{fg:cnst}.b shows the   contour lines for the
lightest scalar-Higgs mass, $\mhone\,$, as  a function of $\mpzero$  and
$\mhpm$.  Notice  that variation in  $\mpzero$ and $\mhpm$ is restricted
to   a  very narrow    region  within which  $\mhone\ge0\,$.  Also,  for
variations  of $\mpzero\,,\mhpm\approxle\order{1}TeV$ it  turns out that
$\mhthree$  remains fairly constant.  Therefore, we  are  free to choose
$\mpzero$ and $\mhpm$ as we    like since $L^+L^-$ production    becomes
insensitive   to $m_{H^0_{1,2}}^{\mbox{}}$ and   $\mpzero$  in the large
$v_3$  limit (which is the   case here).  Furthermore,  the region  over
which $\mpzero$ and $\mhpm$ are allowed  to vary changes insignificantly
for  $v_1/v_2\approxle\order{1}\,$, which   allows  us  to  recant   our
previous    assertion.       Here we   will   consider $\mpzero=200GeV$,
$\mhpm\approx\order{214}GeV$,    $6.7\le v_3/v_2\le 9.1\,$, and $0.02\le
v_1/v_2\le 0.9\,$.

  In general,   the soft-terms should  be evolved   down  from some SUSY
unification scale to give proper masses  to the scalar squarks. However,
as the details of this evolution have not been completely settled, it is
typical to treat these terms as parameters.  Here we will choose $\tilde
M_{\tilde           f}=A_f\equiv\ms\,$,           such              that
$\order{0.4}TeV\approxle\ms\approxle\order{1}TeV\,$.

  The  exotic quark  masses,  the   $m_{q^\prime}$'s,  will be   assumed
degenerate,        such             that        $\order{200}GeV\approxle
m_{q^\prime}\approxle\order{600}GeV\,$.

  Finally,  the $e^{\prime^\pm}$ will be designated  to play the role of
the heavy charged lepton, $L^\pm\,$.

\label{sc:foo}

\subsection{Results}
\begin{figure}[htb]
\mbox{\setlength{\unitlength}{1cm}
\begin{picture}(11.9,5.25)
 \put(-0.15,-0.5){\mbox{\epsfxsize=6.5cm
	\epsffile{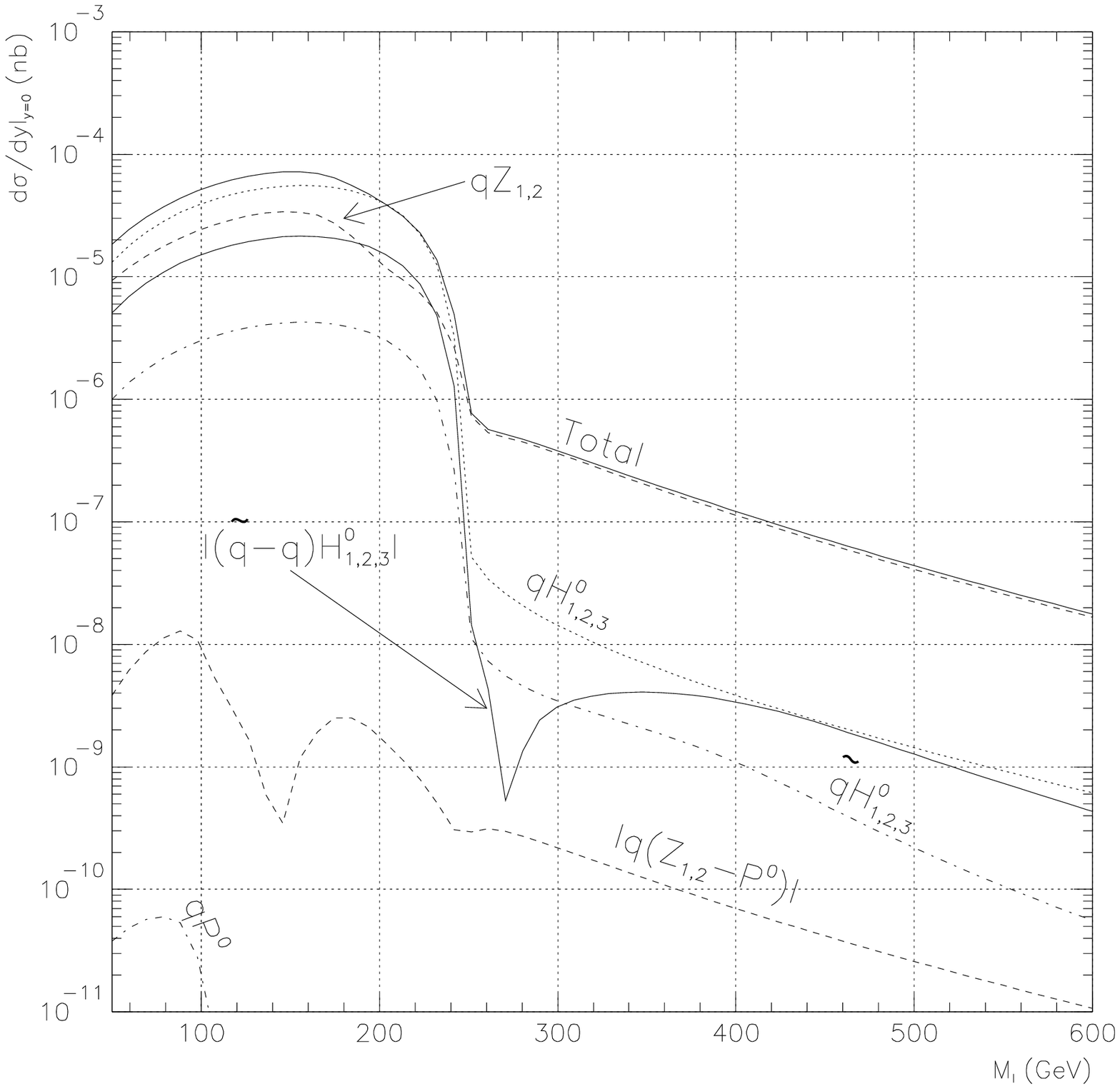}}}
 \put(5.75,-0.5){\mbox{\epsfxsize=6.5cm
	\epsffile{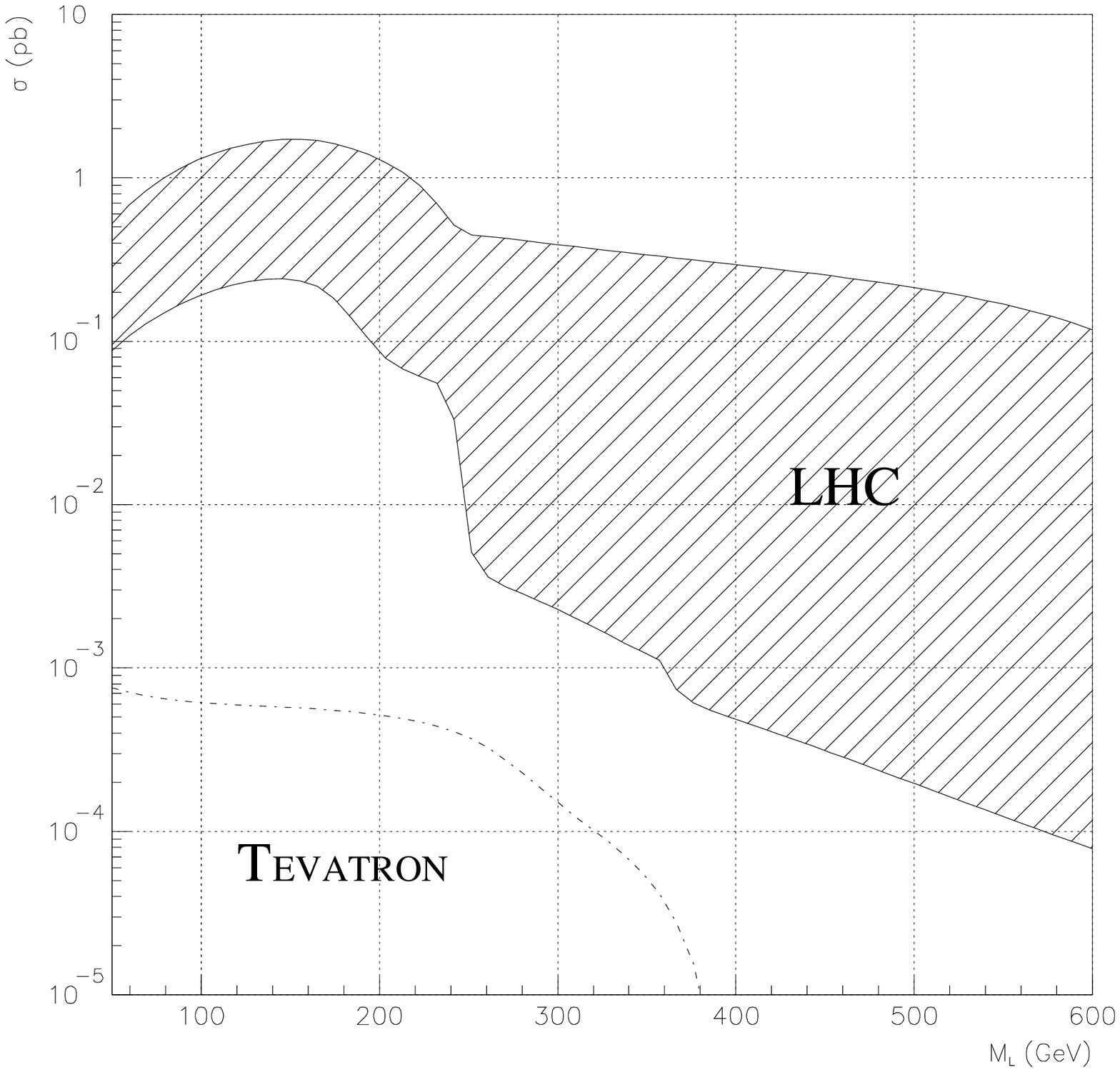}}}
 \put(2.85,0.5){\bf (a)}
 \put(8.85,0.5){\bf (b)}
\end{picture}}
\caption[foo]{Plots of; (a) the rapidity distribution (in $pb$) at $y=0$
     for heavy charged lepton production at $LHC$ as a function of heavy
     lepton  mass,  where     $v_1/v_2=0.02\,$,  $v_3/v_2=6.7\,$,    and
     $\ms=400GeV$.   The  mass spectrum  for the  non-SM  particles are,
     $\mztwo\approx496GeV$                     ($\gztwo\approx20.9GeV$),
     $\mpzero\approx200GeV$                   ($\gpzero\approx16.4GeV$),
     $\mhpm\approx215GeV$,                        $\mhone\approx94.3GeV$
     ($\ghone\approx7.50\times10^{-3}GeV$),        $\mhtwo\approx200GeV$
     ($\ghtwo\approx16.5GeV$),                   $\mhthree\approx495GeV$
     ($\ghthree\approx0.230GeV$), $\mQ=200GeV$.  (b) the results for the
     total $L^+L^-$ production cross-section (in  $nb$) at $LHC$ and the
     $\teva$ as a function of  $\mheavyl\,$.  The hashed region are  the
     $LHC$ results.\hfill\mbox{}}\label{fg:rates}
\end{figure}

  Figure~\ref{fg:rates}.a     shows      the  rapidity     distribution,
$\partial\sigma/\partial y\,$, at  rapidity zero, $y=0\,$, as a function
of   heavy lepton   mass,  $\mheavyl\,$, for  a  typical    set of \Esix
parameters at  $LHC$.\footnote{The rapidity distribution was obtained by
folding  the  parton  level  cross-section   in with the  $DO1.1$  gluon
distribution function.\cite{kn:DO}}
Notice  that the terms   involving $P^0$ are suppressed, as  advertised.
The terms    involving  $H^0_{1,2}$ are also   suppressed,   but are not
explicitly   shown.    The   dramatic drop  in  $\partial\sigma/\partial
y|_{y=0}^{\mbox{}}$ at $\mheavyl\approx250GeV$ corresponds the $Z_2$ and
$H^0_3$ resonance  cut-off's.  The cut-off's vary  with the  $Z_2$ mass,
since $\mhthree\approx\mztwo\approx\order{\gpps    v_3^2}$     for large
$v_3\,$  ($cf$\cite{kn:Gunion}).  Not so  obvious  in this plot, is  the
slight    kink   in       the     $qZ_{1,2}^{\mbox{}}$  curve     around
$\mheavyl\approx200GeV$ ({\it i.e.},  $\approx\mQ$). This corresponds to
the heavy quarks in the loops going off  shell.  The effect becomes more
noticable as the exotic quark masses  are pushed up to $\order{600}GeV$,
because  it   sustains $\partial\sigma/\partial  y|_{y=0}^{\mbox{}}$  at
$\order{10^{-4}}pb$ before gradually starting to drop off.

 Finally,  Figure~\ref{fg:rates}.b  shows the total  $L^+L^-$ production
cross-section at $LHC$ over the \Esix model parameter space specified in
\S~\ref{sc:foo}.  Also shown  are the results  for the $\teva$, with the
parameters given in  figure~\ref{fg:rates}.a$\,$.   Clearly  not  enough
events  are produced at the $\teva$   to make a  search worthwhile, {\it
i.e.}, $\approxle\order{0.1}events/yr\,$.  However,  at  $LHC$ we expect
$\order{10^{4\pm1}}events/yr\,$.  Unfortunately,  this  a factor  of  at
least 10 less than predicted by the $MSSM$.

   The reason for this discrepancy is because $v_3$ is constrained to be
large.  In  this limit  the $L^+L^-$ production  losses  out over $MSSM$
since the number of Higgs propagators has  been reduce from four to one,
whereas $MSSM$ has  three.     Indeed, when $v_3/v_2$ drops   below  the
$\Delta+2\sigma$ contour  line, in figure~\ref{fg:cnst}.a$\,$, the other
Higgses start to   contribute  and $L^+L^-$ production  increases   by a
factor of $\order{10}$, for $\mheavyl\approxle\order{100}GeV\,$.

   It should  also be pointed that,  for  $MSSM$ model  parameters which
yield  a particle spectrum  similar ({\it i.e.}, with comparable masses)
to that of \Esix, a  fair chunk of its  parameter space is eliminated by
unitarity constraints.\cite{kn:Montalvo} In  particular, for the results
shown in figure~\ref{fg:rates}.a$\,$, $MSSM$ production is restricted to
the       region          $\mheavyl\approxle\order{250}GeV\,$,       for
$\mhone\approxge\order{600}GeV$  and    $\tan\beta\approxle\order{5}\,$.
For          more         conservative         $MSSM$        parameters,
$\mheavyl\approxle\order{400}GeV$.

\section{Closing Remarks}
 
  A  simple \Esix  model was constructed  and  used to  compute $L^+L^-$
production    at     high   energy    hadron     colliders.  We   expect
$\order{10^{4\pm1}}events/yr\,$  at $LHC$  and ``zero''  at the $\teva$,
for $50GeV\le\mheavyl\le600GeV$.  The results were a  factor of at least
10 less  than  the $MSSM$  results   due to  the  $CDF$  and $D0\!\!\!/$
soft-limits  on $\mztwo$,\cite{kn:Shochet}  which caused the $H^0_{1,2}$
and $P^0$ contributions to become suppressed.

\section*{Acknowledgments}
This research was funded by $NSERC$ of Canada and $FCAR$ du Quebec.

\end{document}